\documentclass[12pt]{article}
\usepackage{graphicx}
\usepackage{amsfonts,amssymb,amsbsy}
\begin{document}
\title{Quintessential solution of dark matter rotation curves
and its simulation by extra dimensions}
\author{V.V.Kiselev\\[0mm]
\small 
\sl State Research Center "Institute for High Energy Physics" \\[-2mm]
\small {\sl Protvino, Moscow region, 142281 Russia}\\[-1.5mm]
\small \sl Fax: +7-0967-744739,
E-mail: {kiselev@th1.ihep.su} }
\date{}
\maketitle

\vspace*{1mm} \centerline{\includegraphics[width=1.1in]{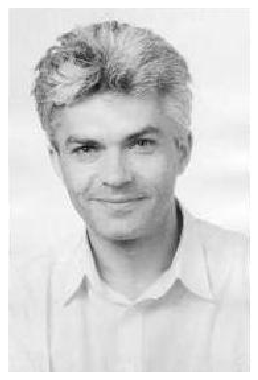}}

\begin{abstract}
On the base of an exact solution for the static spherically
symmetric Einstein equations with the quintessential dark matter,
we explain the asymptotic behavior of rotation curves in spiral
galaxies. The parameter of the quintessence state, i.e. the ratio
of its pressure to the density is tending to $-1/3$. We present an
opportunity to imitate the relevant quintessence by appropriate
scalar fields in the space-time with extra 2 dimensions.
\end{abstract}

\section{Introduction}
\begin{figure}[th]
\begin{center}
\setlength{\unitlength}{0.8mm}
\begin{picture}(85,70)
 \put(0,0){\includegraphics[width=80\unitlength]{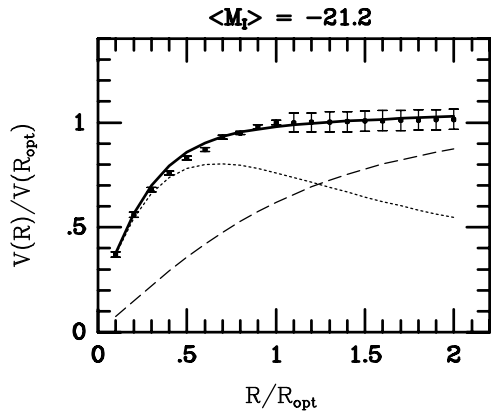}}
\end{picture}
\end{center}
  \caption{The characteristic rotation curve taken from \cite{Olive}.}
  \label{fig-intr}
\end{figure}

The rotation curves in spiral galaxies, i.e. the dependence of
rotation velocity on the distance from the center of galaxy, as
observed astronomically are typically given by the profile
represented in Fig. \ref{fig-intr}.
This picture shows that the contribution determined by the visible
matter (the dotted line) is falling down beyond the optical size
of the galaxy ($R_{\textrm{opt}}$) in agreement with the behavior
expected from the Newton's law for the gravity force of point-like
mass, while in the central disk of galaxy this term is decreased
with the decrease of mass involved in the interaction. The
observed curves prove the presence of dark halo causing the flat,
non-falling character of rotation curves in the asymptotic region
beyond the optical size. The corresponding term shown by the
dashed line begins to dominate at large distances.

The described superposition of two contributions allows a
phenomenological description in terms of universal rotation curves
\cite{univers}. The dark matter review can be found in \cite{rev}.

As for the explanations of the halo dominated contribution
\cite{expl}, we emphasize the attempt of \cite{qdm} to build the
dynamics in terms of scalar fields of the quintessence kind
\cite{quint}, so that the flat rotation curves were found to be
the results of quintessence with the pressure-to-density ratio
close to $-1/3$.

In this paper we apply our recent result on the exact solution of
spherically symmetric static Einstein equations with the perfect
fluid of quintessence \cite{cqg} to the problem of rotation curves
in the asymptotic region of dark-matter-halo dominance. This class
of solutions agrees with the common consideration of static
metrics given in \cite{w}. We find an exact description of
asymptotic behavior in terms of the quintessence and give its
interpretation by the scalar fields in extra 2 dimensions.

\section{Exact results}
In this section, we, first, derive the relation between the metric
components and the rotation velocity. Second, we show how the
quintessential solution reproduces the asymptotic behavior of
rotation curves in the halo dominant region. Third, we explore the
adiabatic approximation in order to describe some variation in the
pre-asymptotic region. Fourth, we give the interpretation of
obtained results in terms of scalar fields in the extra
dimensions.
\subsection{Rotation equations}
We describe the rotation curves in the halo dominated region in
the framework of Hamilton--Jacobi formalism. So, let us consider
the equation for the motion of a test particle with a mass $m$ in
the gravitational field,
\begin{equation}\label{1}
    g^{\mu\nu}\;{\partial_{\mu} S}\,{\partial_{\nu} S} - m^{2} = 0
\end{equation}
with the metric yielding the interval
$$
\textrm{d}s^2 = g_{tt}(r)\,\textrm{d}t^2
-\frac{1}{g_{tt}(r)}\,\textrm{d}r^2 -r^2[\textrm{d}\theta^2+\sin
\theta^2\textrm{d}\phi^2],
$$
which belongs to the class relevant to the problem under interest.
Following the general framework, we write down the solution in the
form, which incorporates two integrals of the motion in the
spherically symmetric static gravitational field,
\begin{equation}\label{2}
    S = -{\cal E}\, t+{\mathfrak M}\,\theta+{\cal S}(r),
\end{equation}
where $\cal E$ and $\mathfrak M$ are the conserved energy and
rotational momentum, respectively. Then, from (\ref{1}) we deduce
\begin{equation}\label{3}
    \left(\frac{\partial {\cal S}}{\partial r}\right)^2 =
    \frac{1}{g_{tt}^2}{\cal E}^2-\frac{1}{g_{tt}}\left(\frac{\mathfrak
    M^2}{r^2}+m^2\right),
\end{equation}
which results in
\begin{equation}\label{4}
    {\cal S} = \int \limits_{r_0}^{r(t)} \textrm{d}r\; \frac{1}{g_{tt}(r)}
    \sqrt{{\cal E}^2-V^2(r)},
\end{equation}
where $V^2$ is an analogue of potential,
$$
V^2(r) = g_{tt}(r)\left(\frac{\mathfrak M^2}{r^2}+m^2\right).
$$
The trajectory is implicitly determined by the equations
\begin{eqnarray}
  \frac{\partial S}{\partial {\cal E}} &=& \textsf{const} = -t
  +\int\limits_{r_0}^{r(t)} \textrm{d}r\; \frac{1}{g_{tt}(r)}
    \frac{\cal E}{\sqrt{{\cal E}^2-V^2(r)}}, \label{p1}\\
  \frac{\partial S}{\partial {\mathfrak M}} &=& \textsf{const} =
  \theta
  -\int\limits_{r_0}^{r(t)} \textrm{d}r\; \frac{1}{r^2}
    \frac{\mathfrak M}{\sqrt{{\cal E}^2-V^2(r)}}.\label{p2}
\end{eqnarray}
Taking the derivative of (\ref{p1}) and (\ref{p2}) with respect to
the time, we get
\begin{eqnarray}
  1 &=& \dot r\; \frac{\cal E}{g_{tt}\sqrt{{\cal E}^2-V^2(r)}}, \\
  \dot \theta &=& \dot r\;\frac{\mathfrak M}{r^2\sqrt{{\cal
  E}^2-V^2(r)}},
\end{eqnarray}
and, hence,
\begin{equation}\label{8}
    {\cal E} = \frac{g_{tt}}{r v}\,{\mathfrak M},
\end{equation}
relating the energy and the rotational momentum, where we have
introduced the velocity
$$
v  \stackrel{\mbox{\tiny def}}{=} r\dot \theta.
$$
The points of return are determined by
$$
\dot r =0, \quad \Rightarrow\quad {\cal E}^2 - V^2 =
0,\quad\Rightarrow\quad {\mathfrak M}^2 = m^2 r^2
\frac{v^2}{g_{tt}-v^2}.
$$
The circular rotation takes place, if two return points coincide
with each other, i.e. we have the stability of zero $\dot r$
condition. Introducing the proper distance $\lambda$ by
$$
\frac{\partial}{\partial\lambda} = \frac{\partial
r}{\partial\lambda}\,\frac{\partial}{\partial r} =
g_{tt}\,\frac{\partial}{\partial r}
$$
we deduce the equation
\begin{equation}\label{10}
    \left(\frac{\partial \cal S}{\partial\lambda}\right)^2 = {\cal
    E}^2- V^2,
\end{equation}
so that the stability of circular motion implies the stability of
potential,
\begin{equation}\label{11}
    \frac{\partial V^2}{\partial r} = 0.
\end{equation}
Then, we get
\begin{equation}\label{12}
    v^2 = \frac{1}{2}\, \frac{\textrm{d}g_{tt}(r)}{\textrm{d}\ln
    r}.
\end{equation}
This result is in agreement with the nonrelativistic
approximation. Indeed, in the Newton's limit, the equality of two
forces, the gravitational attraction and the circular rotation
inertia, gives
$$
F_{NR} = \frac{m}{2}\,\frac{d g_{tt}(r)}{d r} = \frac{m v^2}{r},
$$
which reproduces the exact result of (\ref{12}).

Introducing a re-scaled velocity with respect to the proper time,
$$
{\mathfrak v}^2 = \frac{1}{g_{tt}}\, v^2,
$$
we get the result of \cite{b}
$$
    {\mathfrak v}^2 = \frac{1}{2}\, \frac{\textrm{d}\ln g_{tt}(r)}{\textrm{d}\ln
    r}.
$$
Thus, we have determined the exact profile of circular rotation
curves for the metric under interest.

\subsection{Quintessential solution}
For the spherically symmetric static metric
$$
{\rm d}s^2 = g_{tt}(r)\,{\rm d}t^2+g_{rr}(r)\,{\rm d}r^2-
r^2({\rm d}\theta^2+\sin^2\theta\,{\rm d}\phi^2),
$$
we have found the exact solution \cite{cqg} with
\begin{equation}\label{s1}
    g_{tt} = 1-\sum_n \left( \frac{r_n}{r} \right)^{3w_n+1},
\end{equation}
where $r_n$ are positive constants, and
$$
g_{rr} = -\frac{1}{g_{tt}},
$$
so that the matter averaged over the angles satisfies the perfect
fluid relation between the pressure and energy density with the
state parameter $w_n$,
$$
p_n = w_n \rho_n.
$$
The components of energy-momentum tensor are given by
\begin{eqnarray}\label{t}
    &&{{T^{[n]}}_t}^t = \rho_n(r), \\[2mm]
    &&{{T^{[n]}}_i}^j  =  \rho_n(r)\,3 w_n
    \left[-(1+3\,B_n)\frac{r_i\,r^j}{r_k r^k}+B_n\,
    {\delta_i}^j\right],
\end{eqnarray}
so that the averaging gives
$$
\left\langle {{T^{[n]}}_i}^j\right\rangle = -
p_n(r)\,{\delta_i}^j,
$$
independently of the parameter $B$. However, the Einstein
equations are satisfied at the appropriate value of
\begin{equation}\label{b}
B_n = -\frac{3 w_n+1}{6 w_n},
\end{equation}
which is the only parametrization consistent with the
superposition of various terms. In addition, the above class of
exact solutions covers the characteristic limits such as the case
of a collapsed dust, i.e. a black hole, at $w_0 = 0$, the vacuum
solution, i.e. de Sitter space, at $w_{-1} = -1$, and the
relativistic electromagnetic field of charged black hole at
$w_{em} =1/3$.

The superposition implies that the sum of terms in the time
component of the metric is exactly transformed into the sum of
energy-momentum tensors,
$$
\sum_n \left( \frac{r_n}{r}\right)^{3w_n+1} \Longrightarrow \quad
\sum_{n} {{T^{[n]}}_\mu}^\nu,
$$
so that we can get various exact spherically symmetric static
solutions of Einstein equations by combinations of relevant terms.

Let us consider the limit of quintessence with the state parameter
$w_q= -1/3+\epsilon\to -1/3+0$. Then, the metric component
$$
\tilde g_{tt} = 1
-\frac{\alpha}{\epsilon}\left[\left(\frac{r_q}{r}\right)^\epsilon-1\right]
$$
tends to
\begin{equation}\label{metr}
    \tilde g_{tt} = 1+ \alpha\ln \frac{r}{r_q},
\end{equation}
and in accordance with eq.(\ref{12}) we find the rotation velocity
\begin{equation}\label{va}
    v^2 =\frac{1}{2}\,\alpha,
\end{equation}
describing the asymptotic behavior at large distances, i.e. in the
halo dominated region. Thus, the quintessential solution describes
the asymptotic rotation curves with the metric (\ref{metr}) for
the dark matter.

Numerically, the quantity $\alpha$ is of the order of $10^{-6}$,
which implies that the inner horizon is posed at a distance many
orders of magnitude less than the parameter $r_q$.

\begin{figure}[th]
\begin{center}
\setlength{\unitlength}{1mm}
\begin{picture}(85,50)
 \put(0,0){\includegraphics[width=80\unitlength]{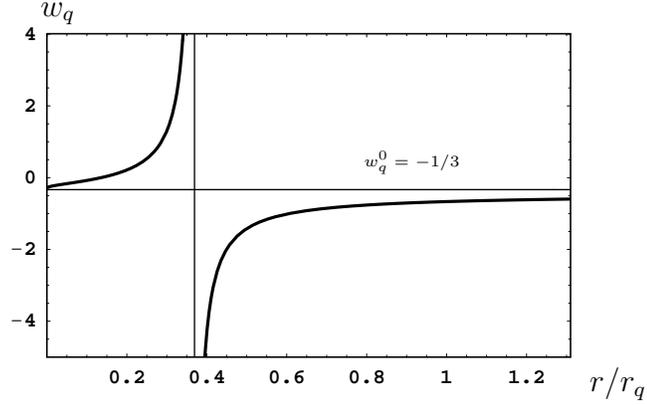}}
 \put(80,0){$r/r_q$}
 \put(7,50){$w_q$}
 \put(50,30){\tiny $w_q^0=-1/3$}
\end{picture}
\end{center}
  \caption{The parameter of state equation for the quintessence.}\label{fig1}
\end{figure}

The energy-momentum tensor for the quintessence is given by the
expression
\begin{eqnarray}\label{tq}
    &&{{T^{[q]}}_t}^t = {{T^{[q]}}_r}^r = \rho_q(r) = -\frac{\alpha}{2 r^2}\,
    \left(\ln\frac{r}{r_q}+1\right) , \\[2mm]
    &&{{T^{[q]}}_\theta}^\theta  = {{T^{[q]}}_\phi}^\phi =  -\frac{\alpha}{4
    r^2}.
    \label{tq'}
\end{eqnarray}
Averaging over the angles results in the parameter of state
equation equal to
\begin{equation}\label{weq}
    w_q = -\frac{1}{3}\left(1+\frac{1}{\ln\frac{r}{r_q}+1}\right),
\end{equation}
which has a singular point at $\ln r/r_q = -1$ (see Fig.
\ref{fig1}).

To the same moment we can isolate two parts in the energy density,
i.e. the logarithmic contribution and the $1/r^2$-term, so that
their state parameters are equal to
$$
w_{\ln} = -\frac{1}{3},\quad w_{1/r^2} = -\frac{2}{3},
$$
respectively. However, this separation is not unique, and any
arbitrary redefinition of $r_q$ parameter will result in the
rearrangement. For example, introducing a large scale $\ln \tilde
r_q/r_q$, we get
$$
\tilde w_{\ln} = -\frac{1}{3},\quad \tilde w_{1/r^2} =
-\frac{1}{3}\,\frac{2+\ln \tilde r_q/r_q}{1+\ln \tilde r_q/r_q}\to
-\frac{1}{3},\quad \textrm{at}\quad \ln \tilde r_q/r_q\to \infty ,
$$
which is the case we have been going to consider with no
singularity of the constant $w$.

The spatial part of energy-momentum tensor for the logarithmic
term is purely radial, while the $1/r^2$-term is tending to the
radial form with the small contribution isotropic over the angles.

Thus, we can draw a conclusion on the dark matter described by the
quintessence with a negative pressure at $w_q=-1/3$ corresponds to
the asymptotically flat rotation curves, so that the exact
solution of static spherically symmetric Einstein equations allows
the superposition of various terms such as the Schwarzschild black
hole surrounded by the quintessence.

\subsection{Adiabatic modification}
Let us consider a phenomenologically motivated variation of the
parameter determining the pre-asymptotic behavior of the term
contributing to the star velocities due to the dark matter halo,
$$
\alpha = \alpha_0\,\frac{r^2}{a^2+r^2},
$$
where the characteristic scale is close to the optical size of the
galaxy, $a\sim R_{\textrm{\textrm{opt}}}$.

An adiabatic variation of parameter $\alpha$ implies that the
variation of energy density caused by a small change of the
parameter versus the distance is much less than the variation due
to the leading dependence on the distance. Therefore, we demand
the condition of adiabatic approximation in the form
$$
\left|\frac{\partial \rho}{\partial
\alpha}\,\frac{\partial\alpha}{\partial r}\right| \ll
\left|\frac{\partial\rho}{\partial r}\right|.
$$
In the metric under study we get
$$
\left|\frac{\partial\ln\alpha}{\partial \ln r}\right|\ll
\left|\frac{\ln \frac{r}{r_q}+1}{\ln \frac{r}{r_q}}\right|.
$$
The phenomenological parametrization of $\alpha$ gives
$$
\frac{2 a^2}{a^2+r^2} \ll \left|\frac{\ln \frac{r}{r_q}+1}{\ln
\frac{r}{r_q}}\right|,
$$
so that the adiabatic approximation is sound at $r\gg a$, while at
$r\sim a$ we deduce the condition
$$
r_q \sim a,
$$
which implies that the quintessence-term parameter $r_q$ is
determined by the optical size of the galaxy, if we desire to
incorporate the observed decrease of the dark matter contribution
into the rotation curves within the description suggested above.
\begin{figure}[th]
\begin{center}
\setlength{\unitlength}{1mm}
\begin{picture}(85,54)
 \put(0,0){\includegraphics[width=80\unitlength]{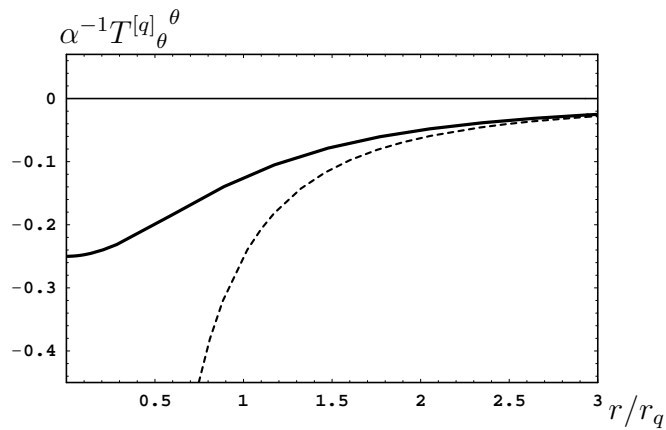}}
 \put(80,0){$r/r_q$}
 \put(7,50){$\alpha^{-1} {{T^{[q]}}_\theta}^\theta$}
\end{picture}
\end{center}
  \caption{The adiabatic regularization of the angle component in
  the energy-momentum tensor of quintessence at $a=r_q$ (the solid curve)
  in comparison with the constant $\alpha$ (the dashed line).}\label{fig-a}
\end{figure}
We present the dependence of angle component for the
energy-momentum tensor of the quintessence under the adiabatic
change of the parameter $\alpha$ in Fig. \ref{fig-a}.

Thus, the decrease of rotation velocity at the distances less than
the optical size of galaxy as caused by the dark matter can be
included in the offered mechanism by the small adiabatic change of
the solution parameter.

\subsection{Interpretation}
The quintessential state with the negative pressure is usually
considered as a perfect fluid approximation of a scalar field with
an appropriate potential. So, let us start with a general
contribution of the scalar field.

A scalar field $\varphi$ with the lagrangian equal to
\begin{equation}\label{lag0}
    {\cal L} = \frac{1}{2}\, g^{\mu\nu}\,\partial_\mu \varphi\,\partial_\nu \varphi
    -V(\varphi),
\end{equation}
generates the energy-momentum tensor
\begin{equation}\label{t0}
    T_{\mu\nu} = \partial_\mu \varphi\,\partial_\nu \varphi
    -g_{\mu\nu}\left[\frac{1}{2}\, g^{\beta\gamma}\,
    \partial_\beta \varphi\,\partial_\gamma \varphi
    -V(\varphi)\right].
\end{equation}
The static field $\varphi(r)$ depending on the distance, gives
\begin{eqnarray}\label{t04}
    &&{T_t}^t = {{T}_\theta}^\theta  = {{T}_\phi}^\phi = \rho(r) = -{\cal L}, \\[2mm]
    &&{{T}_r}^r = {{T}_t}^t - g^{tt}[\partial_r \varphi(r)]^2.
    \label{t04'}
\end{eqnarray}
Eqs. (\ref{t04}) and (\ref{t04'}) compared with eqs. (\ref{tq})
and (\ref{tq'}) imply that the explanation of the quintessential
solution for the rotation curves in the region of dark matter
dominance in terms of the scalar field requires some additional
contributions to the time component of the energy-momentum tensor,
i.e. we should, at least, suppose the presence of a cold dark
matter with the density
$$
\rho_{CDM} = - g^{tt}[\partial_r \varphi(r)]^2,
$$
which is not natural, since the density is negative, and it is
amazingly coherent with the scalar field.

Thus, we do not impose the usual scalar version of quintessence
suitable for the purposes of describing the rotation curves as
caused by the dark matter.

Let us consider the space-time with two extra dimensions, so that
the metric is determined by the interval
\begin{equation}\label{g6}
    \textrm{d}s^2 = g_{tt}(r)\,\textrm{d}t^2
-\frac{1}{g_{tt}(r)}\,\textrm{d}r^2 -r^2[\textrm{d}\theta^2+\sin
\theta^2\textrm{d}\phi^2] + \kappa(y_{-1})\,\textrm{d}y_{-1}^2 -
\kappa(y_{4})\,\textrm{d}y_{4}^2,
\end{equation}
and the 4 dimensional (4D) interval is given by the condition
$$
\textrm{d}y_{-1}^2 = \textrm{d}y_{4}^2.
$$
Introduce two scalar fields by the following definitions:
\begin{equation}\label{-1}
    \bar\varphi = e^{y_{-1}},
\end{equation}
which is the isotropic function, and the triplet
\begin{eqnarray}
  \varphi^{(1)} &=& e^{y_{4}}\sin\theta\,\cos\phi, \\
  \varphi^{(2)} &=& e^{y_{4}}\sin\theta\,\sin\phi, \\
  \varphi^{(3)} &=& e^{y_{4}}\cos\theta,
\end{eqnarray}
depending on the angles.

Then, under the condition of
$$
\kappa(y_{extra}(r)) = r^2,
$$
for the metric components restricted to the 4D world, we find for
the energy-momentum tensors the following expressions:
\begin{eqnarray}\label{t6}
    &&{\bar T_t}^t = {{\bar T}_r}^r = {{\bar T}_\theta}^\theta
    = {{\bar T}_\phi}^\phi = -{\cal L}(\bar\varphi) = -\frac{1}{2 r^2}\,
    e^{2y_{-1}}+\bar V(\bar \varphi),\\[2mm]
    \label{t6'}
    &&{T_t}^t = {{T}_r}^r = -{\cal L}(\varphi^{(i)}) = \frac{3}{2 r^2}\,
    e^{2y_{4}}+V(\varphi^{(i)}),\\[2mm]
    \label{t6''}
    &&{{T}_\theta}^\theta = {{T}_\phi}^\phi = \frac{1}{2 r^2}\,
    e^{2y_{4}}+V(\varphi^{(i)}).
\end{eqnarray}
The 4D space-time is considered at
\begin{equation}\label{=}
    y_{-1} = y_{4},
\end{equation}
so that at
$$
\bar V = V
$$
we can easily get that the scalar fields simulate the
energy-momentum tensor of quintessence, if we put
\begin{eqnarray}\label{v}
  \bar\varphi^2 &=& \left[\varphi^{(i)}\right]^2 = -\frac{\alpha}{4}\,
  \left(2\ln\frac{r}{r_q}+1\right),  \\[2mm]
  \label{v'}
  2 V &=& -\frac{\alpha}{4r_q^2}\,\exp\left[1+\frac{4}{\alpha}\bar\varphi^2
  \right] =-\frac{\alpha}{4r^2}.
\end{eqnarray}
In order to avoid a conflict caused by the sign of
distance-dependent term in eq.(\ref{v}), we make the re-scaling of
the initial lagrangian for the scalar fields by
\begin{equation}\label{re}
    \hat{\cal L} = {\cal L}\cdot {\cal K}(\varphi),
\end{equation}
where, for definiteness, we put
$$
{\cal K} =\frac{1}{\sqrt{\varphi^2}}.
$$
Then, with the same condition on the extra coordinates we derive
\begin{eqnarray}\label{hatv}
  \hat{\bar\varphi}^2 &=& \left[\hat\varphi^{(i)}\right]^2 =
  e^{2y_{-1}(r)} = \frac{\alpha^2}{16}\,
  \left(2\ln\frac{r}{r_q}+1\right)^2,  \\[2mm]
  \label{hatv'}
  2V\cdot{\cal K} &=& -\frac{\alpha}{4r_q^2}\,
  \exp\left[1-\frac{4}{\alpha}\sqrt{\hat{\bar\varphi}^2}\,
  \right] =-\frac{\alpha}{4r^2},\\[2mm]
  {\cal K}^{-1} &=& -\frac{\alpha}{4}\,
  \left(2\ln\frac{r}{r_q}+1\right).
  \label{hatv''}
\end{eqnarray}

We emphasize that the extra dimensional positions of our 4D world,
i.e. the functions $y_{extra}(r)$, are twofold. The same note
should be done on the factor ${\cal K}$ (twofold values of the
square root). The sign of kinetic term for the scalar fields
changes with the sign of $\cal K$. Therefore, we deal with two
branches of the field variation, representing the normal and ghost
phases\footnote{The ghosts have negative sign in front of the
kinetic term in the lagrangian. On a relevance of tachyons in the
modern theory see review in \cite{Gibbons}.}. Nevertheless, the
energy density remains finite.

Thus, we have just shown that the energy-momentum 4D tensor of
quintessence responsible for the flat asymptotic form of rotation
curves in spiral galaxies, i.e. the dark matter contribution, can
be exactly imitated by the scalar fields with 2 extra dimensions.

In this paper we do not investigate the field equations in the
extra dimensions. Note, first, that the corresponding components
of the tensor $R_{-1\,-1} = R_{4\,4} =0$, and they do not
contribute to the 4D Einstein equations through the scalar
curvature. Second, in the extra dimensions we can easily find that
$R_{ab}-1/2 g_{ab} R \neq -2 T_{ab}$, where $\{a,b\}\in \{-1,4\}$,
$R_{ab} = 0$, and $T_{ab}$ is the energy-momentum tensor of scalar
fields under consideration. Of course, we can add two scalar
fields $\bar\chi(y_{-1})$ and $\chi(y_{4})$, so that they have
appropriate potentials producing ${\cal L}(\bar\chi,\chi)=0$, if
the field equations are satisfied, and hence, $\chi$'s do not
contribute to the 4D Einstein equations, while their derivatives
are adjusted in order to make the Einstein equations valid in the
extra dimensions. In that case the extra-dimensional components of
energy-momentum tensor are proportional to the metric, i.e. the
situation in extra dimensions looks like the vacuum solution with
the curvature depending on the 3D distance as a parameter. So, the
scalar field equations are topics of separate investigations.
Thus, we claim only that the extra-dimensional scalar fields with
appropriate exponential potentials simulate the quintessential
solution for the rotation curves.

By the way, once we have encountered the problem of ghosts in the
treatment with extra dimensions, we have to note that equations
equivalent to (\ref{hatv})--(\ref{hatv''}) can be replicated in
the 4D space-time, so that the only difference is the overall
negative sign of the lagrangian for the triplet field\footnote{We
have also suggested that the squares of scalar fields depend on
the radius but the extra coordinates.}.

The change of normal phase to the ghost for the scalar field is
spectacular, since it is closely related to the variation of sign
for the energy density. Both these facts are irrelevant in the
case of adiabatic growth of the parameter $\alpha$ at $\ln r/r_q
\approx 0$. The ghost phase is essential in the asymptotic region
of large distances, if we do believe in the constant velocity of
rotation in infinity. On another side, the negative value of
kinetic energy is familiar from the quantum mechanics if the
particle enters the classically forbidden region under the
potential barrier.

\section{Conclusion}
In this paper we have found a quintessential solution for the
problem on the asymptotic behavior of the rotation curves in
spiral galaxies in the region of dark-matter-halo dominance. The
explanation is constructed on the base of new class of metrics,
describing the perfect fluid with a negative pressure in the
static spherically symmetric gravitational field. This class
satisfies the principle of superposition for various kind of
matter contributions, and, hence, it does not destroys the
Schwarzschild metric by adding some amount of exotic or ordinary
matter.

The quintessence with the state parameter $w_q=-1/3$ exactly
results in the flat limit of rotation curves. Its energy-momentum
tensor is simulated by scalar fields in extra dimensions with
appropriate exponential potentials.

The author thanks Prof. R.Dzhelyadin for the possibility to
collaborate with the LHCB group at CERN, where this work has been
done. I am grateful to Prof. S.S.Gershtein for useful discussions
and valuable remarks, and to members of Russian team in LHCB for a
kind hospitality.

This work is supported in part by the Russian Foundation for Basic
Research, grants 01-02-99315, 01-02-16585, and 00-15-96645.

\end{document}